\documentclass[runningheads]{llncs}
\usepackage{graphicx}
\begin{document}


\title{Usage of single-camera video recording to measure sea surface roughness with machine learning methods}

\titlerunning{Preprint. Proc. of CCIE 2023. To be submitted to LNEE}   
%
\author{Mikhail B. Salin\inst{1}\orcidID{0000-0001-8260-5422} \and
Artem V. Vitalsky \inst{1,2} }
%
\authorrunning{}    
%
\institute{Gaponov-Grekhov Institute of Applied Physics of the Russian Academy of Sciences (IAP RAS), Nizhny Novgorod, 603950, Russia \\
\email{mikesalin@ipfran.ru} \and
Lobachevsky State University of Nizhny Novgorod (UNN), Nizhny Novgorod, 603950, Russia}
\maketitle              
\begin{abstract}
Photometry is a convenient operational method for monitoring such dynamically evolving phenomena as wind waves. Nowadays machine learning allows one to avoid explicit derivation of the solution to the problem, describing all the instructions for transforming the input data into the final result. Instead, an algorithm is used to independently find solutions through the integrated use of statistical data, from which patterns are derived, on the basis of which forecasts are made. An example of a problem for which a regular solution has traditionally been applied is the prediction of wave height from the input brightness of a water surface. The task is complicated by the multitude of possible physical models and the need to apply calibration coefficients. In this paper, we solve the problem of how, basing on the obtained brightness values and the corresponding heights, to train the neural network to further predict the heights from the incoming brightness values with the greatest accuracy.

\keywords{geophysical signal processing \and image processing \and optical measurement of surface roughness \and video processing \and wind-generated waves}
\end{abstract}
\section{Introduction}
The task of studying the characteristics of waves in the sea is of great interest in oceanology. At present, wave monitoring is carried out visually, as well as with the help of instruments, mounted on ships, platforms and buoys, as well as from satellites \cite{yuan2016method}. Since sea waves in daylight are clearly visible to the naked eye, the development of a method for measuring their parameters from photo and video records began with the advent of the first suitable computing devices and image acquisition tools \cite{gotwols1980optical}. (See also modern reviews, written by the researches, who started to work those days \cite{munk2009inconvenient,titov2021}.) Nowadays, even mass-market and semi-professional cameras have high resolution and frame rate, so they are becoming an affordable and easy-to-use tool for oceanographic research \cite{salin2015combined,salin2015combined2}.

The difficulty lies in the fact that the measurement of wave heights from a video image is an indirect measurement method. Images of a wave with the same amplitude and frequency may be different depending on the shooting conditions, lighting and environment. In order to find a short and understandable review on this issue engineers might have a look on a papers, written by those, who solve the opposite problem: generating realistic images of the sea for video games \cite{abrosimov1999,ma2016real}.

In some cases, stereo imaging is used to reliably measure the wave height \cite{benetazzo2012offshore}, i.e computer vision methods are applied similar to human binocular vision. Such measuring systems require more time to prepare for work, and also have a limitation in ratio between the maximum and minimum measured wave. Sometimes the polarization of light reflected from the sea surface is used as a feature to train the model \cite{mitnik2009remote,ginio2023efficient}.

Within the framework of the method developed by the authors of this article, one camera can be used to determine the shape and characteristics of the sea surface due to the fact that the brightness of the water is corrected with its actual elevation above the average water level. Previously, a mathematical method was proposed for applying calibration data to an image \cite{salin2015combined}. Modern machine learning methods make it possible to determine a much more complex functional relationship between quantities than a simple correlation \cite{GAD2021127}. In this work, the model dependences of the wave surface brightness on its height at the initial stage were set analytically in accordance with their physical interpretation - to test the algorithm. The conducted research can be divided into stages:
\begin{itemize}
    \item  Generation of synthetic data based on the physical model, i.e. images simulating the Stokes wave
    \item Pre-processing of real data
    \item Training and testing the neural network model
\end{itemize}

At the initial stage, an approach was used to process more realistic generated images simulating Stokes waves, changing some parameters, such as camera position, shooting time and angle, frame rate per second. The final step is prediction on real data. In our work, we used the convolutional type of neural network architecture as a basis, which has proven itself well for working with images.

At the time of writing, there are a number of articles on the application of deep learning methods to problem solving in geosciences. As one example, \cite{iafolla2022sea} deals with solving the problem of restoring gaps in buoy datasets using microseismic measurements and machine learning methods.

\section{Problem statement}
\begin{figure}
\includegraphics[width=\textwidth]{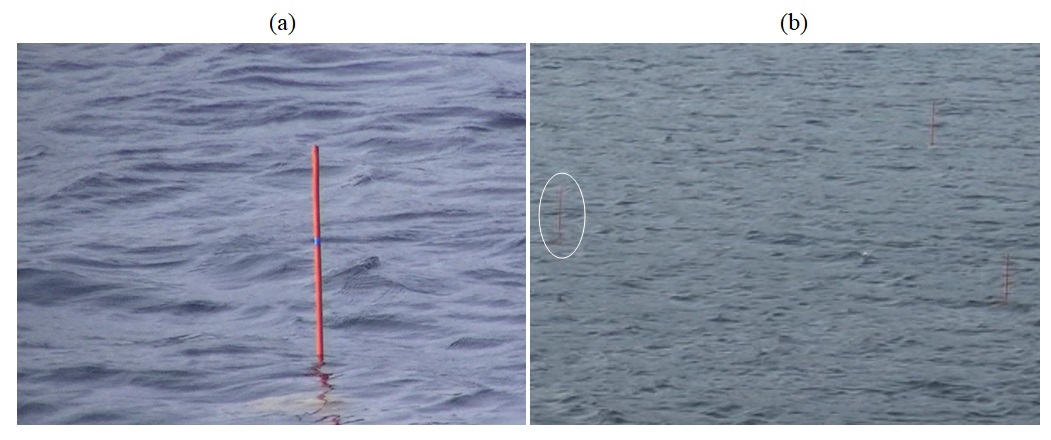}
\caption{Samples of (a) zoom-in and (b) zoom-out frames of surface waves
with a setup of buoys} \label{fig:video_sample}
\end{figure}

A video image of the wave surface is given (like Fig.~\ref{fig:video_sample}b). It is necessary to find a functional relationship between the waveform and its brightness in the image. To solve the problem, a "ground truth" datasets are provided, which are samples of height oscillations in limited number of points, where buoys are located (see  Fig.~\ref{fig:video_sample}a as a zoom-in of panel b). In the subsequent processing we shall draw conclusions about the rest of the areas.

Modern wave-rider buoys measure waves directly from the fluctuations in the water at the point where they are located. Additionally, the buoy can be equipped with a camera and then it will transmit the brightness of the surface in the form of an image in a certain neighborhood around it.

In this work, at the initial stage, we used synthetic data generated in accordance with a physical model to test the algorithm. Based on the obtained brightness values and the corresponding heights, we train the neural network to further predict the heights from the incoming brightness values. 

\begin{figure}
\includegraphics[width=\textwidth]{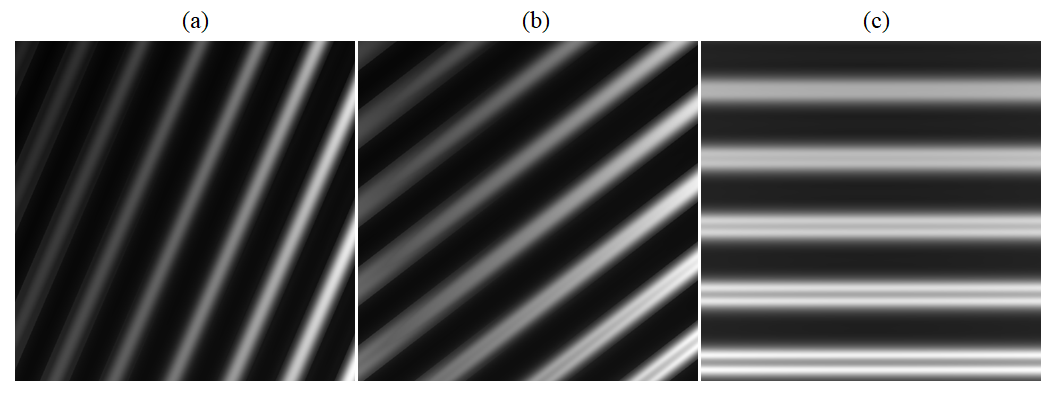}
\caption{Frames generated to further process OpenCV and train the neural network} \label{fig:test_sample}
\end{figure}

Images simulating Stokes waves were generated (see Fig.~\ref{fig:test_sample}). The parameters, listed below, were changed to take different values in different data sets. A neural network was trained on the obtained images. Each generated image of the water surface corresponds to a matrix, the dimension of which coincides with the size of the image, which stores the height value corresponding to each pixel in the image. When creating images, one can take into account such characteristics as:
\begin{itemize}
    \item Camera position (Height, horizontal distance of the camera to the view area and width of the view area)
    \item Shooting time (that influence the sun position and sky radiance)
    \item Shooting angle
    \item Frames per second
\end{itemize}
And also it is possible to adjust the parameters of the waves themselves.

\begin{figure}
\centering
\includegraphics[width=6cm]{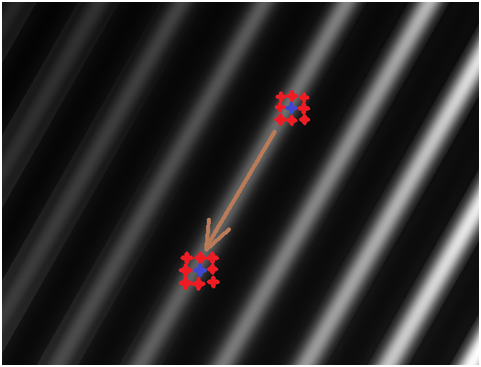}
\caption{Comments on image processing} \label{fig:test_markup}
\end{figure}

After the images of the waves are generated, they are processed, namely, the number of pixels {\it N} and the corresponding height values are selected and then fed to the training of the neural network. Initial theoretical models suggest that the brightness of a surface can be related to its slope \cite{salin2015combined} or even curvature. Therefore, it is fundamentally necessary to predict the height at the blue point by supplying a set of pixels in its vicinity to the model input, as shown in Fig.~\ref{fig:test_markup}. In this case, the neural network will have the physical ability to extract the necessary features. Roughly speaking, the neural network, due to its layers, will be able to build a spatial filter with a certain mask

Fig.~\ref{fig:test_markup} shows the approach to image processing, namely:
with a system of {\it N} pixels, we pass in the direction of the arrow and collect the RGB values in these coordinates (these N pixels are marked with crosses, and the direction of their movement is marked with a beige arrow). A blue cross marks the place where the height is restored.

\section{Calculation algorithm}
\begin{figure}
\includegraphics[width=\textwidth]{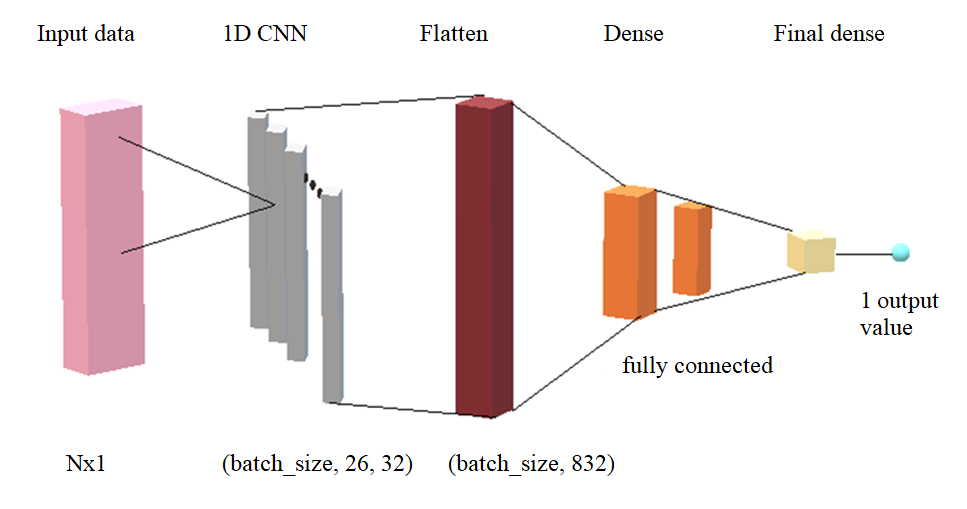}
\caption{Layout of the network } \label{fig:cnn}
\end{figure}

We used convolution neural network (CNN) as the architecture of the neural network since it is very good at identifying simple patterns in the data being used, which in turn will serve to form more complex patterns in the next layers. The layout is shown on Fig.~\ref{fig:cnn}.

\paragraph{Input data}: The data has been processed in such a way that the input is an Nx1 matrix. Actually the camera provides frames like Fig.~\ref{fig:test_markup}, and we split in into small fragments. Namely at one time we pick RGB-to-grayscale values in each pixel marked by a red cross on that figure, and that is the input Nx1 set.
\paragraph{The first layer of 1D CNN}: In the starting layer after the input layer, a filter (feature extractor) of height 2 is determined. Defining one single filter in the first layer allows the network model to notice the dependence trend on one single feature, therefore, in the network used, 32 filters are determined by sampling, which allows you to train 32 different functions on the initial layer of the model. The output of the first inner layer of the used neural network model is a 26x32 neural matrix. Each column of the output matrix contains the weights of one individual filter. With a certain kernel size and given the length of the input matrix, each filter will contain 26 weights.
\paragraph{Layer 2 Flatten}: Used to convert incoming data to a lower dimension. In our case, the input dimension layer (batch\_size, 26,32) is "flattened" to the output dimension (batch\_size, 832).
\paragraph{Layer 3 Dense}: Implements the operation: $output = activation(dot(input, kernel) + bias)$, where activation is the elementwise activation function passed as the activation argument, kernel is the weight matrix created by the layer, and bias is the displacement vector created by the layer. Thus, the output of this layer remains 64 values.
\paragraph{Final layer Dense}: Performs the same functions, but has 1 output value.

\begin{figure}
\includegraphics[width=\textwidth]{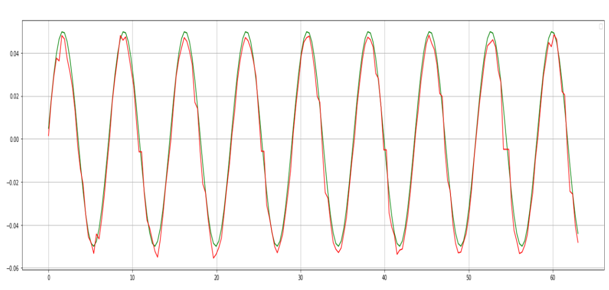}
\caption{The result of prediction by a neural network on a synthetic dataset } \label{fig:model_pred}
\end{figure}

Below is the result of predicting the synthetic data after training the model on the same way synthesized test dataset (Fig.~\ref{fig:model_pred}). These results encourage us to go on with processing real data.

\section{Real data}
The final third stage of the current work is the processing, training and testing of the constructed neural network model on real data. In the study, we used 2 videos of the water surface. One is needed to extract the brightness characteristics of the water surface of the water (Fig.~\ref{fig:video_sample}b), and the second is to obtain the value of the wave height (Fig.~\ref{fig:video_sample}a). These 2 videos should have been synchronized with each other with an accuracy of around 0.01 s, cut into frames with a shooting frequency of 25 frames per second.

\begin{figure}
\includegraphics[width=10cm]{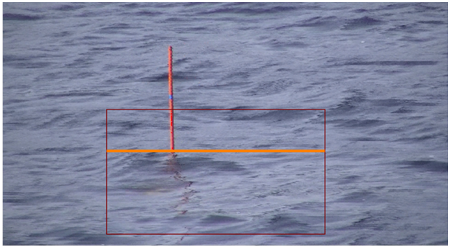}
\caption{An example of extracting the value of the wave magnitude on a random frame} \label{fig:buoy_markup}
\end{figure}

As a result of the actions above, we have 2 packs of frames for obtaining a test and training set: one is for extracting the brightness values in the area of processing , the second is the one from which the height values will be obtained. We exploit a special image processing technique \cite{borodina2010estimation} to extract height values for frames like Fig.~\ref{fig:buoy_markup}, using the buoy that is just a stick as a reference. Any other kind of a sensor that can be synchronized with video will work in that place. However we had such kind of instrumentation that time. A small number of parameters is set in the software to ensure that the horizontal line is fixed clearly under the pole and measures the magnitude of the wave during the automatic processing. Among these parameters are:
\begin{itemize}
\item Field (the area on the image is selected, inside which the stick will be located during the processing of all frames);
\item Color (by clicking on the stick, the color in RGB format will be automatically selected, which will be oriented to the horizontal line that fixes the wave height);
\item Thresholds (sensitivity to the color of the stick and minimum number of pixels to identify the lowest point of the stick);
\item Connectivity (sensitivity to the associated group of pixels with the specified color, used to increase the accuracy of the recognition).
\end{itemize}

The result of the study is highly dependent on the quality of data preparation. Data preparation consisted of the following steps:
\begin{itemize}
\item Removal or smoothing of uncertain data;
\item Scaling values, very often effectively affects the convergence rate when training a neural network, for example, in the preferable range of values from 0 to 1 or from -1 to 1;
\item Implementation of a clear correspondence between input and output data.
\end{itemize}

\begin{figure}
\centering
\includegraphics[width=9cm]{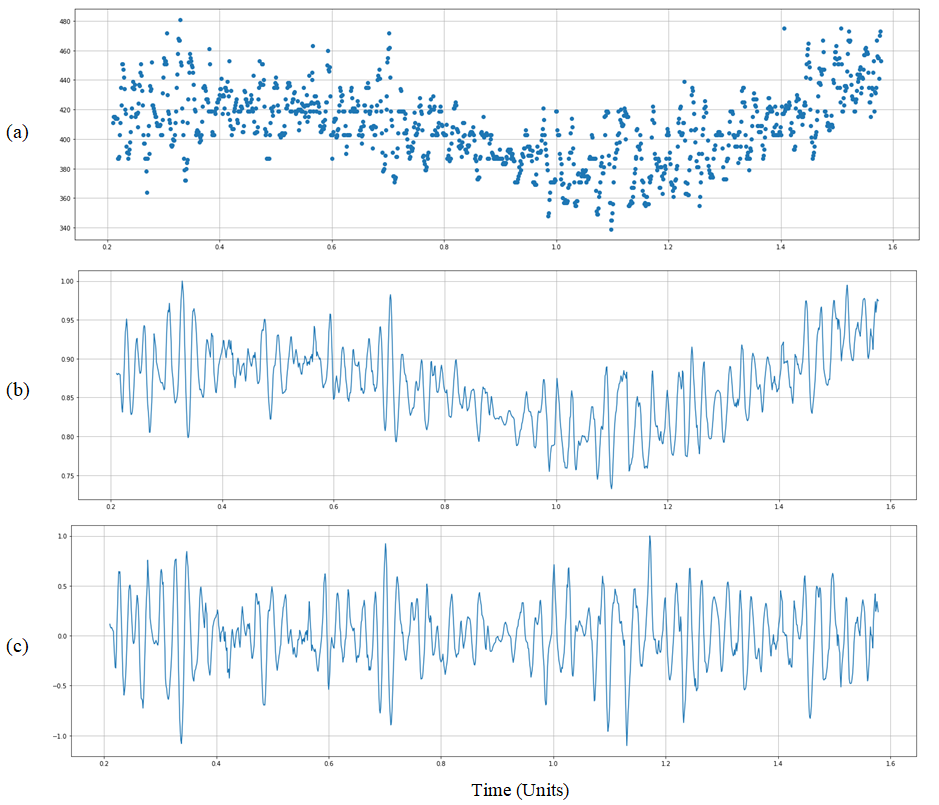}
\caption{Height, i.e. the data to be predicted: (a) initial data, (b) after moving average, (c) after high pass filter} \label{fig:smooth}
\end{figure}

Poor data quality causes difficulties in processing, and in addition to that makes it difficult to determine the trend and the relationship between input and output values. The best way in this case is not to delete such data instances, but to bring them to an adequate form. In our case, the original data looked like Fig.\ref{fig:smooth}a.

Next, to bring the data set in order, a moving average algorithm was used. After calling this function, our data takes on a more understandable form for capturing dependencies (see Fig.\ref{fig:smooth}b). Next, a high-pass filter was used in order to eliminate a slow trend. Data takes the final form in the range of values from -1 to 1 (see Fig.\ref{fig:smooth}c).

Scaling  of input data has a positive effect on the operation of the neural network. This is due to the fact that after applying the scaling method, the input values are located in a single range of values and have a single order for the entire sample. By defining each instance of the training and test sets in one fairly narrow range of values, we will be sure that each instance will equally likely affect the bias of the weights in the course of work. In addition to all of the above, in neural network architectures that use sigmoid functions in the activation function, this avoids oversaturation of neurons, which means that neurons remain operating.

We split the data into control and training samples. The size of the control sample from the entire data set was 10\%. The test data must satisfy 2 criterias. It should be large enough to provide statistics. It should be formed according to the same principle as the training set, but should not be part of the training subset. 

We exploit the routine, explained above for image processing. Recall a set of red and blue crosses, designating input and output locations on the wave, shown on Fig.~\ref{fig:test_markup}. We select the similar set of points on the real data frame (Fig.~\ref{fig:video_sample}) so that the blue center is near the stick lowest end.

\begin{figure}
\includegraphics[width=\textwidth]{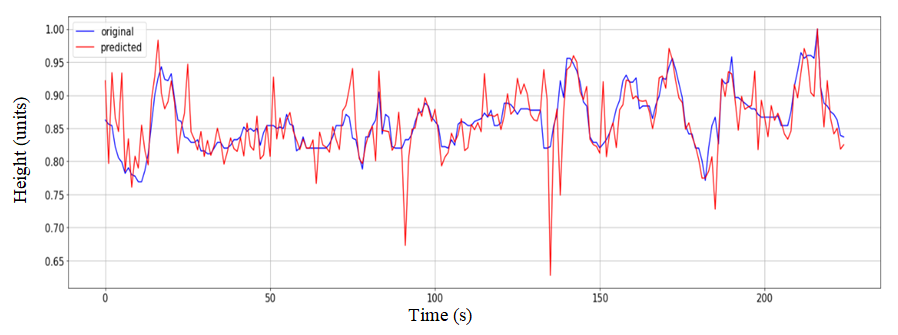}
\caption{Height, i.e. the data to be predicted: (a) initial data, (b) after moving average, (c) after high pass filter} \label{fig:the_result}
\end{figure}

As for the results of real data processing by the neural network on, Fig.~\ref{fig:the_result} helps us to compare the true and predicted height values in arbitrary units of measurement. It can be seen that our model manages to notice the trend quite well and shows confident results on the test set. Weights were well-chosen in the learning process. Probably, the prediction accuracy can be increased by extending the training dataset or by improving the way of selecting input data.

\section{Conclusion}
In this paper, methods, algorithms for machine learning and data analysis were applied to the problem of estimating the height of waves from a video image. The original processing algorithm was written in Python using the Keras and PyTorch libraries. Data preprocessing was carried out and the convolution neural network was trained. This study results show that the model predicts the wave height from the brightness of the water surface with acceptable accuracy, and in general, this approach is promising for further development.

\subsubsection{Acknowledgment}
We are grateful to Dmitry Razumov for providing synthetic data for testing and we also would like to thank Alexander Ponomaremko for valuable discussions. 

 M.Salin's research, namely data acquisition, was supported by the Russian Science Foundation, grant No. 20-77-10081. A.Vitalsky's research on machine learning-based processing was supported by the State Contract with the Ministry of Education and Science of the Russian Federation, ref. No. 0030-2021-0017.

%
%
%
\bibliographystyle{splncs04}
\bibliography{video-waves}

%
%
%
%
\end{document}